\documentclass[12pt]{iopart}

\usepackage[pdftex]{graphicx}
\usepackage[pdftex]{color}

\usepackage{iopams}

\expandafter\let\csname equation*\endcsname\relax 
\expandafter\let\csname endequation*\endcsname\relax

\usepackage{amsmath}
\usepackage{amssymb}
\usepackage{mathrsfs}
\usepackage{citesort}
\usepackage{setspace}

\newcommand{\beq}{\begin{equation}}
    \newcommand{\eeq}{\end{equation}}
    \newcommand{\beqa}{\begin{eqnarray}}
    \newcommand{\eeqa}{\end{eqnarray}}

    \def\mrm{\mathrm}
    \def\mbf{\mathbf}

    \def\max{\mathrm{max}}
    \def\min{\mathrm{min}}

    \def\ww{\mrm{w}}
    \def\WW{\mrm{W}}

    \def\RR{\mathbb{R}}
    
    \def\xx{\mbf{x}}
    \def\XX{\mbf{X}}
    \def\tx{\tilde{x}}
    \def\txx{\tilde{\mathbf{x}}}

    \def\yy{\mbf{x}^\prime}

    \def\ff{\mbf{f}}
    \def\FF{\mbf{F}}
    \def\tF{\tilde{F}}
    \def\tFF{\tilde{\mathbf{F}}}

    \def\xxi{\boldsymbol \xi}

    \def\Nmb{S_\mrm{mb}}

\usepackage{color}

\definecolor{blue}{rgb}{0.0, 0.0, 1.0}
\definecolor{darkgreen}{rgb}{0.0, 0.4, 0.0}
\definecolor{purple}{rgb}{0.50, 0.0, 0.50}

\begin{document}

\title
[Dimensional Reduction of Dynamical Systems by Machine Learning]
{Dimensional Reduction of Dynamical Systems by Machine Learning: 
Automatic Generation of the Optimum Extensive Variables and Their Time-Evolution Map}

\author{Tomoaki Nogawa}

\address{$^1$ Department of Medicine, Faculty of Medicine, Toho University, 
5-21-16, Omori-Nishi, Ota-ku, Tokyo, 143-8540, Japan}
\ead{nogawa@med.toho-u.ac.jp}
\vspace{10pt}
\begin{indented}
\item[] May 2023
\end{indented}

\begin{abstract}
A framework is proposed to generate a phenomenological model that extracts the essence of a dynamical system (DS) with large degrees of freedom using machine learning. For a given microscopic DS, the optimum transformation to a small number of macroscopic variables, which is expected to be extensive, and the rule of time evolution that the variables obey are simultaneously identified. The utility of this method is demonstrated through its application to the nonequilibrium relaxation of the three-state Potts model. 
\end{abstract}


\section{Introduction}

In the studies of the dynamics of systems with large degrees of freedom, 
we often use macroscopic phenomenological models with small degrees of freedom, 
e.~g., the Lorenz equation for atmospheric variability \cite{Lorenz1963}. 
By adopting such a model, we can then use various mathematical techniques developed in the field of nonlinear dynamics to analyze it.
Such a reduced model is usually introduced with drastic approximation 
owing to the intuition and insight of researchers. 
Although it is an orthodox task for statistical physicists 
to derive such a reduced model ab initio from a well-established microscopic model, 
it is possible in highly limited cases such as all-to-all coupling systems. 
This paper proposes a generic framework 
to generate a macroscopic dynamical system (DS) 
for a given microscopic DS with help from machine learning. 
One of the key features of modern machine learning, 
represented by deep neural networks \cite{Carleo2019}, 
is the automatic extraction of feature amounts of the input data. 
In a similar manner, 
we try to find suitable variables to describe the dynamics of a many-body system. 
The DS obeyed by the obtained variables is also an outcome of learning.

The most standard strategy for the dimensional reduction of data is 
mode decomposition, which includes principal component analysis (PCA). 
For many-body dynamics, proper orthogonal decomposition \cite{Berkooz1993} 
has been developed in the field of fluid dynamics. 
Recently, dynamic mode decomposition \cite{Rowley2009,Schmid2010}, which is based on the theory of Koopman operator \cite{Koopman1931}, has attracted attention.
A similar idea is utilized for stochastic processes, 
e.g., the dynamics of protein molecules, 
as Markov state model \cite{Molgedey1994,Perez-Hernandez2013,Wu2017}. 
In mode decomposition analysis, 
the reduced variables are given as 
the coefficients of the eigenvectors of a kind of covariance matrix. 
Usually, the only way to speculate the meaning of the variables is to visualize the eigenvector.
This works well when the eigenvectors can be related to some macroscopic inhomogeneity  
such as obstacles for fluid and the higher-order structure of protein molecules. 
In general, this is not easy because the dimensions of the eigenvector are as large as those of the original data.
The method presented here employs a different type of reduced variable. 
That is extensive variables 
such as internal energy or the volume of a specific local state. 
They are given by spatial integration of local quantities,  
which can be characterized by relatively small number of parameters. 
It works even though the system does not have any characteristic spatial structure. 
In this paper, We apply the method to the Potts model with translation symmetry.

The rest of this paper is organized as follows. 
The basic framework of machine learning to reduce the dimension of DSs is introduced in section~2 
and computational implementation is explained in section~3.  
In section~4, an explanation of how to prepare the data for the demonstration is provided, 
and in section~5, the results of the learning are presented. 
Finally, section~6 provides concluding remarks and highlights future research possibilities.

\section{Formulation}

\subsection{what to learn}

Let us start with a microscopic DS 
\beq
\xx \in \RR^N \mapsto \yy = \ff(\xx) \in \RR^N, \  N \gg 1. 
\label{eq:microDS}
\eeq
The goal is to obtain a macroscopic DS in the form 
\beq
\XX(\yy) = \FF\left(\XX(\xx) \right) \in \RR^n, \ n \ll N, 
\label{eq:macroDS}
\eeq
where $\XX$ is projection onto a macroscopic variable and $\FF$ is time evolution map. 
The task for our machine learning is to find $\XX$ and $\FF$ that satisfy Eq.~\eqref{eq:macroDS} 
from some data points in form $(\xx, \yy)$ that satisfy Eq.~\eqref{eq:microDS}. 
We do not necessarily need to know $\ff$ 
and can use the data sampled from observational time series. 
If the time interval $\Delta t$ between $\xx$ and $\yy$ is not uniform among data, 
we should replace Eq.\eqref{eq:macroDS} with 
$\XX(\yy) = \XX(\xx) + \FF\left(\XX(\xx) \right) \Delta t$.

If $\XX$ is given, the task is a regression of $\FF$, that is, supervised learning. 
Many studies have been conducted on this type of DS regression \cite{Schmidt2009,Brunton2016}.
If $\XX$ is not given, it is not a popular problem. 
It is, however, similar to the finite-size scaling for critical phenomena \cite{Harada2011,Harada2015}, 
where we simultaneously seek for how to scale the variables 
and what equation the scaled variables satisfy.

In the actual machine learning, we fix $n$ and 
seek for $\XX$ and $\FF$ that minimize the loss functional (LF)  
\beq
L_n[\XX,\FF] := \frac{1}{n} \  \overline{ \left| \FF\left(\XX(\xx) \right) - \XX(\yy) \right|^2 }. 
\eeq
Hereafter, the overline denotes the average taken over data points. 
If $L_n[\XX^*,\FF^*]$ equals zero, $(\XX^*, \FF^*)$ gives an exactly closed DS. 
Otherwise, it gives an approximated formula, whose precision is evaluated by $L_n$. 
We naively expect that $\min_{\XX,\FF} L_n$ decrease with $n$. 
The $n$-dependence would inform us a kind of the complexity of the system. 
A likely scenarios  are, for example, 
$\min L_n=0 \iff n \ge n_1$, $\min L_n \propto e^{-n/n_2}$, and so on.

\subsection{orthonormal condition}

There exists indefiniteness in $\XX$; 
If $(\XX^*, \FF^*)$ satisfies Eq.~\eqref{eq:macroDS}, 
$(\mathbf{G} \! \circ \! \XX^*, \mathbf{G} \! \circ \! \FF^* \! \circ \! \mathbf{G}^{-1} )$ does too 
with arbitrary function $\mathbf{G}:\RR^n \to \RR^n$ that has an inverse function. 
To avoid the indefiniteness, we need to impose some restrictions on $\XX$.

First, we impose orthonormal condition on $\XX(\yy)$ in data space as  
\beq
\overline{ X_m(\yy) X_l(\yy) } = \delta_{ml} 
\quad \forall m, l \in \{1, \dots, n\}.
\label{eq:orthonormal}
\eeq
Here, we suppose $\overline{ \XX(\yy) } = \mbf{0}$.  
We impose the condition not on $\XX(\xx)$ but on $\XX(\yy)$ 
because the latter corresponds to the LF more directly. 
Normalization excludes the trivial solution $(\XX,\FF)=(\mbf{0},\mbf{0})$. 
Orthogonalization avoids the solutions where $X_m \approx X_l \quad \forall m,l$. 
This solution is favored 
because the number of the objective variables, namely $\{ X_m(\yy) \}_{m=1}^n$, is effectively reduced to one.  
On the other hand, the number of the explanatory variables, $\{ X_m(\xx) \}_{m=1}^n$, is not reduced 
because the difference $X_m-X_l$ is allowed to be blown up in $\FF(\XX)$.

\subsection{extensiveness}

As the second restriction on the macroscopic variables, 
we suppose that $\XX$ is given by the summation of a local $b$-body function 
$\boldsymbol\xi: \RR^b \to \RR^n$ as  
\beq
\XX(\xx) = \frac{1}{N} \sum_{i=1}^{N} \ \xxi (x_{\nu_{i1}}, \dots, x_{\nu_{ib}}). 
\label{eq:extensive}
\eeq
Here $\{ \nu_{ij} \}_{j=1}^b$ indicates the block that includes $i$ 
and its neighbors in some sense. 
This makes $\XX$, precisely $N \XX$, an extensive variable. 
Even if the microscopic dynamics is stochastic, 
we expect that the time evolution of extensive variables is deterministic for $N \to \infty$. 
The definition of the block would be customized for the DS to analyze. 
If the microscopic variables are embedded on vertices in a graph, neighbors are naturally determined by edges. 
In the case of particles in continuous space, 
we need to define edges between particles by, 
for example, introducing cut-off distance or making Voronoi tessellation.

Hereafter, we consider the case that the microscopic variables are discrete and bounded 
as $x_i \in \{0,1,\dots, q-1 \} := \Xi$. 
The set of the states of a block is the Cartesian power $\Xi^b$. 
Thus, arbitrary local functions can be expressed as 
$\xxi(x_{\nu_{i1}}, \dots, x_{\nu_{ib}}) = \sum_{k \in \Xi^b} \mbf{w}_k \delta_{k_i k}$, 
where $k_i$ is identical to $(x_{\nu_{i1}}, \dots, x_{\nu_{ib}})$.  
Substitution of this into Eq.~\eqref{eq:extensive} leads to 
\beqa
\XX(\xx) = \sum_{k \in \Xi^b} \mbf{w}_k \tx_k(\xx)
, \quad  
\tx_k(\xx) := \frac{1}{N} \sum_{i=1}^{N} \delta_{k_i k}. 
\label{eq:linear_combination}
\eeqa
Here $\tx_k$ means the fraction of the blocks that take the configuration $k$. 
Hereafter we note 
\beq
\XX(\xx) = \ww \txx(\xx), 
\ \ww := (\mbf{w}_0, \dots, \mbf{w}_{q^b-1}) \in \RR^{n \times q^b}, \quad 
\txx(\xx) := ^t \!\! \left( \tx_0(\xx), \dots, \tx_{q^b-1}(\xx) \right) \in \RR^{q^b},  
\eeq
where $t$ on the left shoulder denotes the transposition. 
In the case that $\xx$ is a continuous variable, we have to consider different formulations. 
A naive solution is to employ polynomial bases 
such as $\tx_k(\xx) = N^{-1} \sum_i x_{\nu_{i1}}^{\,a_{k1}} \dots x_{\nu_{ib}}^{\,a_{kb}}$.

As shown in Eq.~\eqref{eq:linear_combination}, 
$\XX$ is expressed by linear combination of $\{\tx_k(\xx) \}_{k\in \Xi^b}$ 
and we can employ $\left( \txx(\xx), \txx(\yy) \right)$ 
as a data point instead of $\left( \xx, \yy \right)$. 
This saves the computational cost in the machine learning when $q^b \ll N$. 
All bases $\{\tx_k\}_{k\in\Xi^b}$ are not linear-independent, 
for instance, we have the relation $\sum_{k\in \Xi^b} \tx_k = 1$.  
Let the number of the linear-independent bases be $N_b$.  
We need $N_b \ge n$ to satisfy the orthonormal condition Eq.~\eqref{eq:orthonormal}.

\begin{figure}[t]
     \includegraphics[trim=0 160 160 10,scale=0.45,clip]{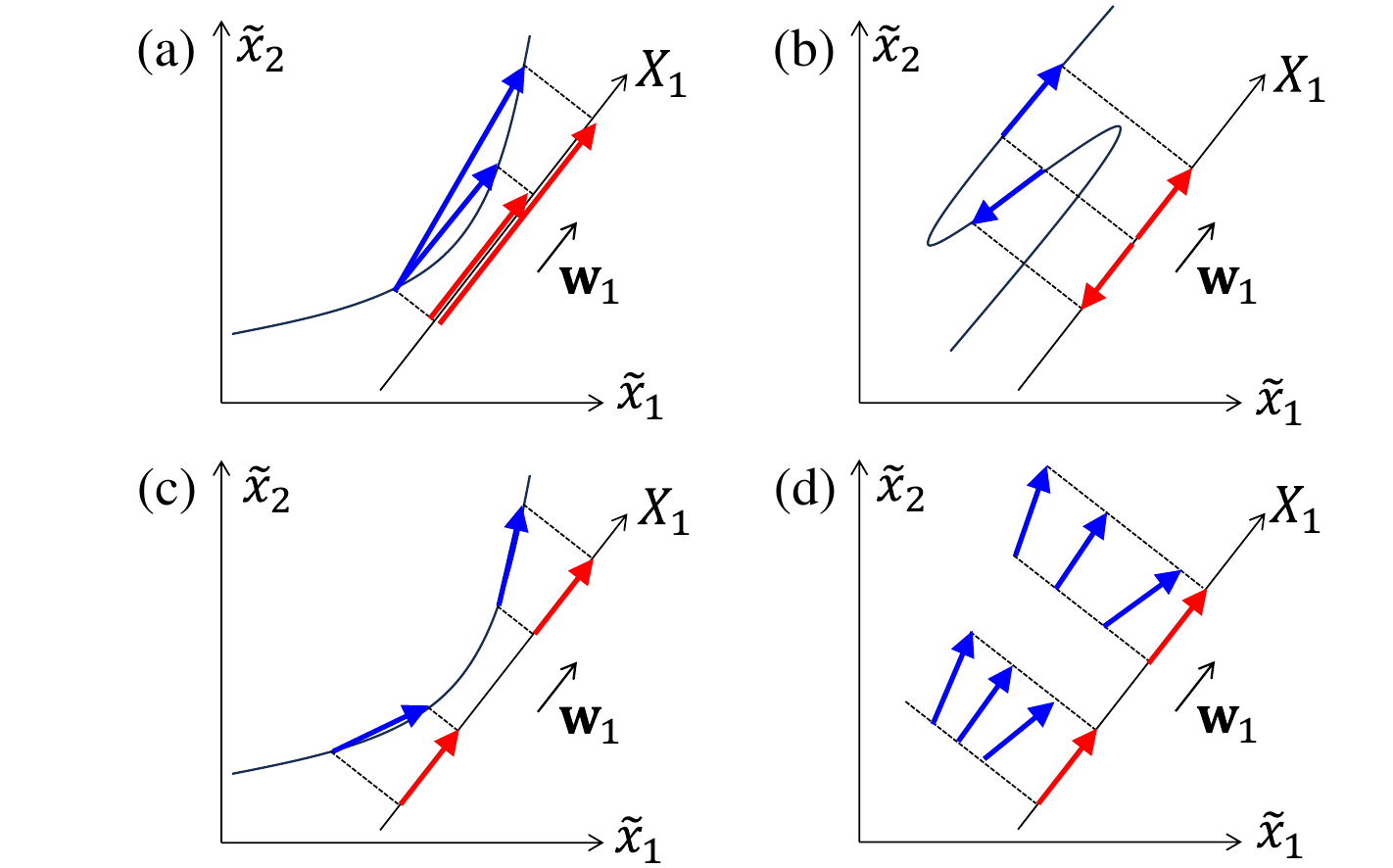}

    \caption{
        Schematic illustration of the projection of the displacement of $\txx \in \RR^2$ onto $X_1$-axis. 
        \label{fig:schematics}
    }
\end{figure}

Let us summarize how dimensional reduction is achieved in the present formulation. 
It is done by two steps as $\xx \in \RR^N (\mathrm{or} \  \Xi^N) \mapsto \txx \in  \RR^{N_b}$
and $\txx \in \RR^{N_b} \mapsto \XX = \ww \txx \in \RR^n$.  
The latter is a projection onto a $n$-dimensional hyperplane in $\RR^{N_b}$ 
that is parallel to $\mathbf{w}_m$ $\forall m \in \{1, \dots, n\}$. 
We say that the DS of $\XX$ is closed if the map $\XX(\xx) \mapsto \XX(\xx^\prime)$ is unique. 
Its sufficient conditions are (i) the map $\txx(\xx) \mapsto \txx(\xx^\prime)$ is unique 
and (ii) $\txx \mapsto \ww \txx$ is in one-to-one correspondence. 
Figure~\ref{fig:schematics} illustrates several scenarios for $N_b=2$ and $n=1$. 
Panel~(a) shows a failed case, where condition (i) is not satisfied. 
The non-uniqueness of $\txx(\xx) \mapsto \txx(\xx^\prime)$ might be solved by increasing $b$. 
Panel~(b) also shows a failed case, where condition (ii) is not satisfied. 
To satisfy it, the manifold where $\txx$ is embedded should be {\it directed}. 
Panel~(c) shows a successful case where the two conditions are satisfied. 
Panel~(d) also shows a successful case, 
where we obtain closed DS even though condition (ii) is not satisfied.  
Note that condition (ii) is not a sufficient condition.  
It is, however, likely that $\txx(\xx)$ is embedded in a directed manifold 
whose dimension is equal to or less than $n$ when $\XX(\xx) \mapsto \XX(\xx^\prime)$ is unique.

\subsection{dropout}

Although $L_n^*$ decreases by increasing $b$, 
it becomes more difficult to interpret the meaning of $\XX(=\ww \txx)$ 
because the number of the elements of $\ww$ expands as $q^b$. 
It is desirable that $\ww$ is sparse; most of the elements are zero. 
For this purpose, we enforce {\it dropout} of the bases in the following manner. 
We allocate a subset $\Xi_m^b \subseteq \Xi^b$ for $m \in \{ 1, \dots, n \}$ 
and set $w_{mk}=0$ unless $k \in \Xi_m^b$.  
To satisfy the orthonormal condition, we need to suppose $|\Xi_m^b| \ge m$. 
(The reason for this condition will be explained in Sec.~\ref{sec:orthonormalization}.)  
We remark that dropout eliminates the rotational indefiniteness of $\XX$.

\subsection{model of macroscopic dynamical system}

On the function form of $\FF$, 
we simply suppose an $n$-variable polynomial up to the $p$-th order as 
\beqa
\FF(\XX) = \WW \tFF(\XX), \ \WW \in \RR^{n \times N_p}, \ 
N_p := \frac{(p+n)!}{p!n!}, 
\nonumber \\
\tF_l(\XX) = \prod_{m=1}^n {X_m}^{a_{l m}}, \ 
a_{l m} \in \{0,1,\dots,p \} , \ \sum_{m=1}^n a_{l m} \le p. 
\label{eq:polynomial_of_F}
\eeqa
Eventually, the LF is expressed as a function of the weight 
$(\ww, \WW) \in \RR^{n\times(q^b+N_p)}$. 

In the learning process, we try to reduce the LF by tuning $\ww$ and $\WW$ alternately. 
When $\ww$ is fixed, the task is linear regression, and $\WW$ is optimized by solving 
\beq
\overline{ \tFF\left(\XX(\xx)\right) ^t\! \tFF\left( \XX(\xx)\right) } \WW 
= \overline{ \tFF\left(\XX(\xx)\right) ^t\! \XX(\yy)}.
\label{eq:regression}
\eeq 
When $\WW$ is fixed, we iteratively update $\ww$ via gradient descent.  
Further details on the learning process are explained in the next section.

\section{Numerical implementation}

\subsection{learning procedure}
\label{sec:apdx1}

Before starting the learning, we generate a dataset containing  
$2^{16}$ samples of $\left( \txx(\xx), \txx(\yy) \right)$. 
We can make data with various $b$ from single $\left( \xx, \yy \right)$. 
As a pretreatment, we subtract $\overline{\txx(\yy)}$ from both $\txx(\xx)$ and $\txx(\yy)$, 
so that $\overline{\XX(\yy)}=\mathbf{0}$ for arbitrary $\ww$ without a bias term. 
If the Pearson correlation coefficient between bases $\tx_k$ and $\tx_l$ exceeds 0.999,  
we merge these bases into $\tx_k+\tx_l$. 
By using this dataset, we perform the following computation {\it almost} independently on 32 threads.

At the beginning of training, initialization is done as follows. 
\begin{enumerate}
\item Set $\ww$ as $w_{mk} \sim U(-0.5, 0.5)$.  
\item Set $\WW$ by solving Eq.~\eqref{eq:regression}. 
\item Choose the size of minibatch $\Nmb$ randomly from $\{2^{8}, 2^{9}, 2^{10}, 2^{11}, 2^{12}\}$. 
\item Choose the number of redundant components $\delta$ randomly from $\{0, 1, \dots, 8\}$ 
and make the subsets $\{ \Xi_{m}^b \}_{m=1}^n$ s.t. $|\Xi_m| \ge m$ and $\sum_{m=1}^n (|\Xi_{m}^b|-m) = \delta$. 
\end{enumerate}
Here, $U(a,b)$ means the uniform distribution between $a$ and $b$. 
If dropout is not done, step (4) is skipped.

We regard $2^{20}/\Nmb$ updates of $\ww$ as an epoch of training, 
whose CPU-time overhead rarely depends on $\Nmb$. 
We repeated it $2^{12}$ times at most. 
At the beginning of an epoch, 
we choose $S_\max(=2^{14})$ samples from the dataset for training data   
and test data, respectively. 
Here, samples are randomly chosen by avoiding duplication. 
The training data are divided into $S_\max/\Nmb$ minibatches. 
Then, we iterate the following. 
\begin{enumerate}
\item Choose one minibatch randomly. 
\item Update $\ww$ by the Mini-batch gradient descent \cite{Bottou2010} with ADAM \cite{Kingma2015}.
\item Update $\WW$ for every 4 updates of $\ww$.
\item Calculate $L_n(\ww,\WW)$ by using {\it test} data for every $2^{16}/\Nmb$ update of $\ww$ 
and renew the record of $L_n^*$ if $L_n<L_n^*$.   
\end{enumerate}
At the end of an epoch, we perform the initialization in the threads that satisfy the abandonment condition:     
(i) the momentary $L_n$ is larger than the median among the threads
and (ii) $L_n^*$ has not been renewed over the last two epochs. 
Such threads are not likely to reach the optimum solution. 
This operation is the sole exception of the independence among the threads.

\subsection{Orthonormalization}
\label{sec:orthonormalization}

After every update of $\ww$, including random initialization, 
we enforce the orthonormalization as follows. 
First, we orthogonalize $\{ X_m(\yy) \}_{m=1}^{n}$ by changing $\mrm{w}$ as 
\beq
w_{mk} \to w_{mk} - \sum_{l=1}^{m-1} \frac{ \overline{ X_l X_m } }{\overline{ {X_l}^2 }}  w_{lk} 
\left( \iff 
X_{m} \to X_{m} - \sum_{l=1}^{m-1} \frac{ \overline{ X_l X_m } }{\overline{ {X_l}^2 }}  X_{l}  
\right). 
\label{eq:orthogonalization}
\eeq
Here, we subtract the components parallel to $X_l$ for all $l<m$ from $X_m$. 
This is done sequentially in the fixed order: $m=2, 3, \dots, n$. 
The average $\overline{\tx_{k_1} \tx_{k_2} }$, which appears in the calculation   
\beq
\overline{X_l X_m} = \sum_{k_1\in \Xi^b} \sum_{k_2\in \Xi^b} 
w_{l k_1} w_{m k_2} \overline{\tx_{k_1} \tx_{k_2} },
\eeq
is taken not over the samples in the minibatch but over the whole training data.  
Since it needs to be calculated only once for an epoch, the computational overhead is low. 
Secondly, we perform normalization as 
$w_{mk} \to \left( \overline{ {X_m}^2 } \right)^{-1/2} w_{mk}$. 
Even if the record-law loss function is obtained, 
we do not accept it unless both of the two conditions:  
\beq
\underset{m,l\, |\, m\ne l}{\max} \left| \overline{ X_m X_l } \right|^2 < 10^{-4}
\quad \mrm{and} \quad 
\underset{m}{\max} \left| \overline{{X_m}^2}-1.0 \right|^2 < 10^{-3}  
\label{eq:cond_orthonormalization}
\eeq
are not satisfied with the test data.

In cases involving dropout, we cannot treat the term with $w_{lk}$ in Eq.~\eqref{eq:orthogonalization}
unless $X_m$ includes the base $\tx_k$. 
We introduce $Z_l := \sum_{k \in \Xi_m^b} v_{lk} \tx_k$ 
as a substitute of $X_l= \sum_{k \in \Xi_l^b} w_{lk} \tx_k$ in Eq.~\eqref{eq:orthogonalization}. 
Here, $\{v_{lk} \}$ is determined by solving the simultaneous equations 
$\partial \overline{ |Z_l - X_l|^2 }/\partial v_{lk} = 0 \quad \forall k \in \Xi_m^b$. 
Since we make $X_m$ orthogonal to $X_l \  \forall \ l < m$, we need $|\Xi_m| \ge m$ in general. 
Consequently, the minimum number of the nonzero elements in $\ww$ equals $\sum_{m=1}^n m=n(n+1)/2$.

\section{Data}

\subsection{Metropolis-Hastings dynamics of three-state Potts model}

In this paper, we show the application of the method 
proposed above to the three-state Potts model. 
The internal energy of the system is given by 
$E(\xx) := -\sum_{ (i,j) \in \mrm{n.\,n.} } \delta_{x_i x_j}$, 
where $x_i \in \{0, 1 ,2\}$ is the spin variable at the lattice point $i$ 
in a $(L\times L \times L)$-cubic lattice 
and ``n.\,n.'' means the set of all nearest-neighbor pairs of the lattice points 
under the periodic boundary condition. 

For the microscopic dynamics, we employ the Metropolis-Hastings update \cite{Metropolis53,Hastings70}. 
Starting with a certain initial state, which will be explained later, 
we iterate the following elementary trial. 
First, we make a candidate new state $\xx_2$ from existing state $\xx_1$ 
by choosing a lattice point $i$ at random and replacing $x_i$ with one of the other two spin states with equal probability.
We accept the change to $\xx_2$ with probability 
$\min \left\{ 1, \exp \left[ K \left\{ E(\xx_1)-E(\xx_2) \right\} \right] \right\}$. 
Here $K$ is the ratio of the coupling constant to the temperature. 
We regard that time $t$ increases by $1/N$ for each elementary trial.  
From a single time sequence, we sample one data point 
by recording $\xx$ at time $t$ and $\yy$ at $t+\Delta t$, where $t \sim U(t_\min, t_\max)$. 
In this work, we set $t_\min=1.0, t_\max=16.0$ and $\Delta t = 1.0$.

\subsection{initial condition}

For each time sequence, we set the initial state by assigning $x_i$ values randomly and independently.
We make the probability distribution of the spin state differs among samples.  
Let the probability to assign $x_i=\sigma$ be $p_\sigma$.   
We consider three types of distribution.  
In initial condition (IC) I,  $p_\sigma = r_\sigma/(r_0 + r_1 + r_2)$ 
where $r_0, r_1, r_2 \sim U(0,1)$ independently. 
In IC II, we set $r_1=r_2$, which leads to $p_1=p_2$. 
In IC III, we set $r_0=r_1=r_2$, which leads to $p_0=p_1=p_2$.

\subsection{blocks}

As the block of neighbors, namely $\{ \nu_{ij} \}$, 
we adopt the sequential straight chain with length $b-1$ 
that starts at lattice point $i$ and elongates on the [1,0,0]-direction of the cubic lattice. 
By this definition, arbitrary bases for $b-1$ can be expressed by those for $b$ 
since we have 
\beq
\tx_{x_1 x_2 \cdots x_{b-1} } 
= \sum_{x_{b} \in \Xi} \tx_{x_1 x_2 \cdots x_{b-1} x_{b}} 
= \sum_{x_{b} \in \Xi} \tx_{x_{b} x_1 x_2 \cdots x_{b-1}} 
\quad \forall x_1 x_2 \cdots x_{b-1} \in \Xi^{b-1}.
\label{eq:b_and_b-1}
\eeq 
Here the suffixes of $\tx$ are not products but $b$-digit $q$-nary integer. 
Therefore, the minimum of the loss function never increases with $b$. 
Since we have $q^{b-1}$ relations among bases, that is, Eq.~\eqref{eq:b_and_b-1},  
$N_b$ equals $q^b - q^{b-1}$ at most, which holds even for $b=1$.    
(The sum rule $\sum_{k\in \Xi^b} \tx_k = 1$ is not independent of these equations.)

For $b=1$, we have $\txx=(\tx_0, \tx_1, \tx_2)$.  
Each base is identical to the magnetization $M_\sigma = (3 \tx_\sigma - 1)/2$. 
In the following, we do not consider constant bias and coefficient in the expression of the macroscopic variables.
Since $\tx_0+\tx_1+\tx_2=1$, the number of independent magnetization equals two. 
For $b=2$, we have $\txx=(\tx_{00},\tx_{01}, \dots ,\tx_{21},\tx_{22})$,
where $\tx_{\sigma_1 \sigma_2}=\tx_{\sigma_2 \sigma_1}$ holds.   
This can express internal energy $E=\tx_{01}+\tx_{12}+\tx_{20}$ 
as well as $M_\sigma = \tx_{\sigma0} + \tx_{\sigma1} + \tx_{\sigma2}$.

\section{Results}

Now, we show the results of the machine learning. 
First, we fix $K = 1.20K_c$, where $K_c \approx 0.5506$ \cite{JANKE1997679} is the threshold  
above which the system has a nonzero spontaneous magnetization 
in equilibrium ($t \to \infty$, $N \to \infty$). 
In the latter part, we consider the data with various $K$. 
Dropout is not considered before Sec.~\ref{subsec:dropout}. 

In the following, the optimum value of the LF is denoted by $L_n^*$. 
Precisely, it is the average of the four best results of 32 nearly independent learning processes.  
Note that $L_n^*$ depends not only on $n$ but also on $b$, $p$, $N$, IC, and so on. 
If not declared, we set $p=5$ and $N=256^3=2^{24}$.

\begin{figure}[t]
    \includegraphics[trim=20 60 420 20,scale=0.35,clip]{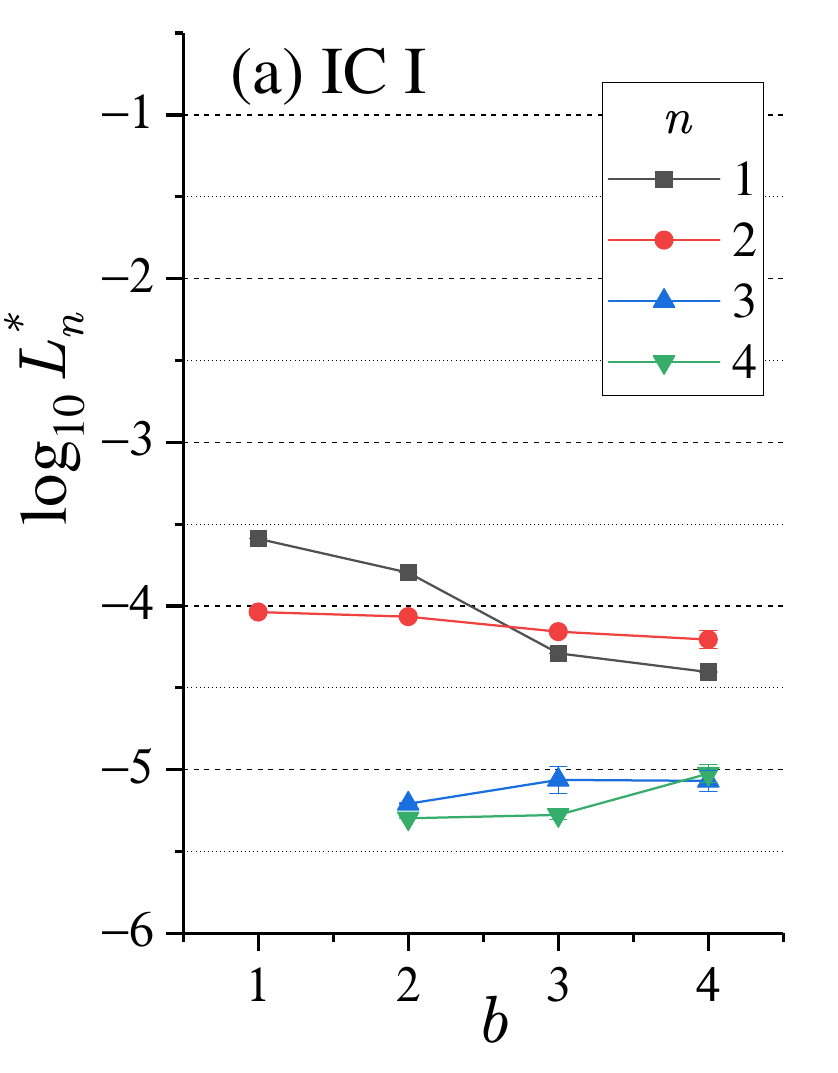}
    \includegraphics[trim=20 60 420 20,scale=0.35,clip]{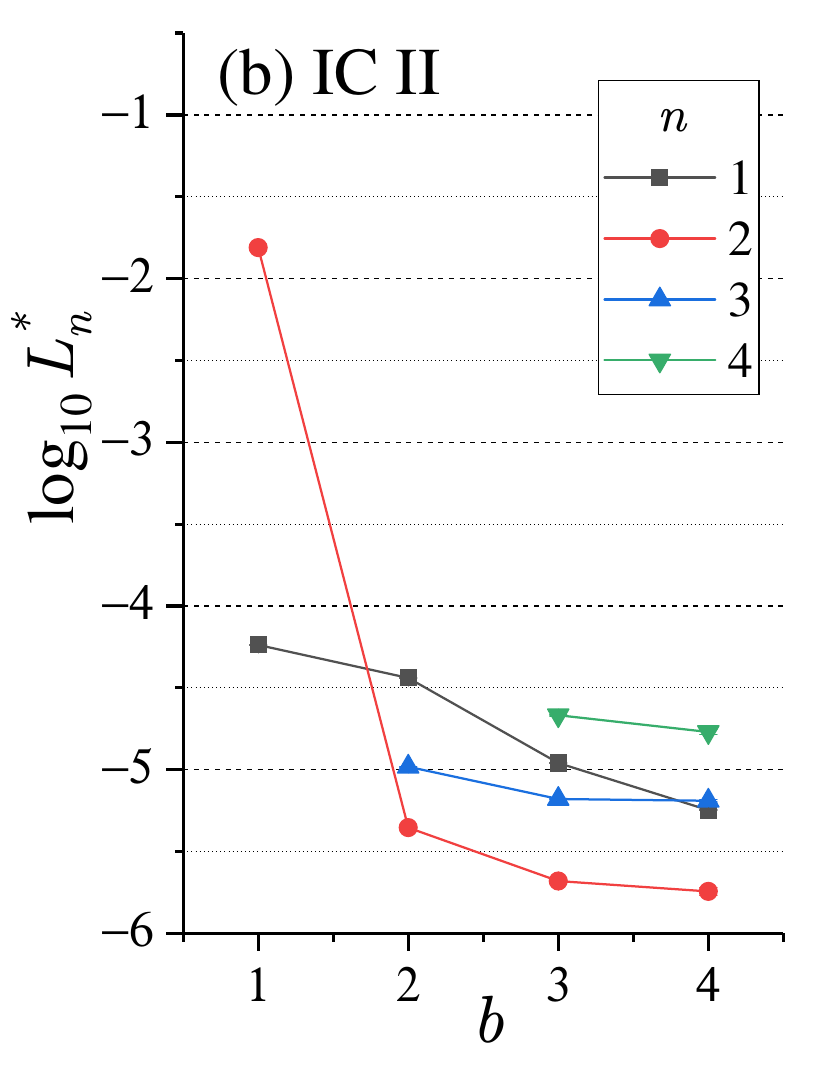}
    \includegraphics[trim=20 60 420 20,scale=0.35,clip]{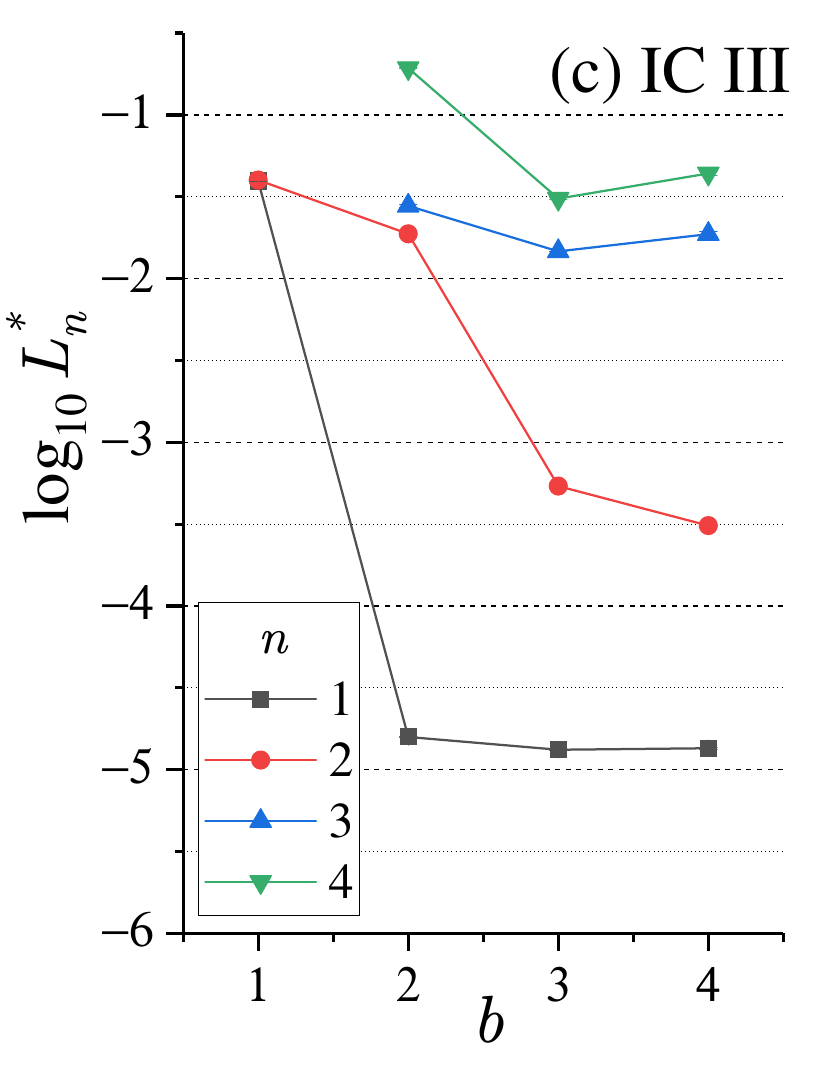}
    \caption{
        The optimum value of the loss function obtained by the learning 
is plotted as a function of the block length.
Panel (a), (b), and (c) correspond to IC I, IC II, and IC III, respectively. 
The error bars, which are too small to see, indicate the root mean square deviation 
of the four best results. 
        \label{fig:L-b}
    }
\end{figure}

\subsection{initial condition I}

Under IC I, the symmetry among the three spin states is fully broken in general, 
and the majority state at $t=0$ remains the majority for $t>0$.   
Figure~\ref{fig:L-b}(a) shows the $b$-dependence of $L_n^*$ for each $n$. 
For $n\ge3$, $L_n^*$ increases with $b$ due to the incompleteness of learning. 
Anyway, the $b$-dependence of $L_n^*$ is not large. 
For fixed $b$, $L_n^*$ decreases with $n$. 
Great improvement is observed between $n=2$ and $n=3$.

\begin{figure}[t]
\begin{center}
    \includegraphics[trim=0 50 280 20,scale=0.35,clip]{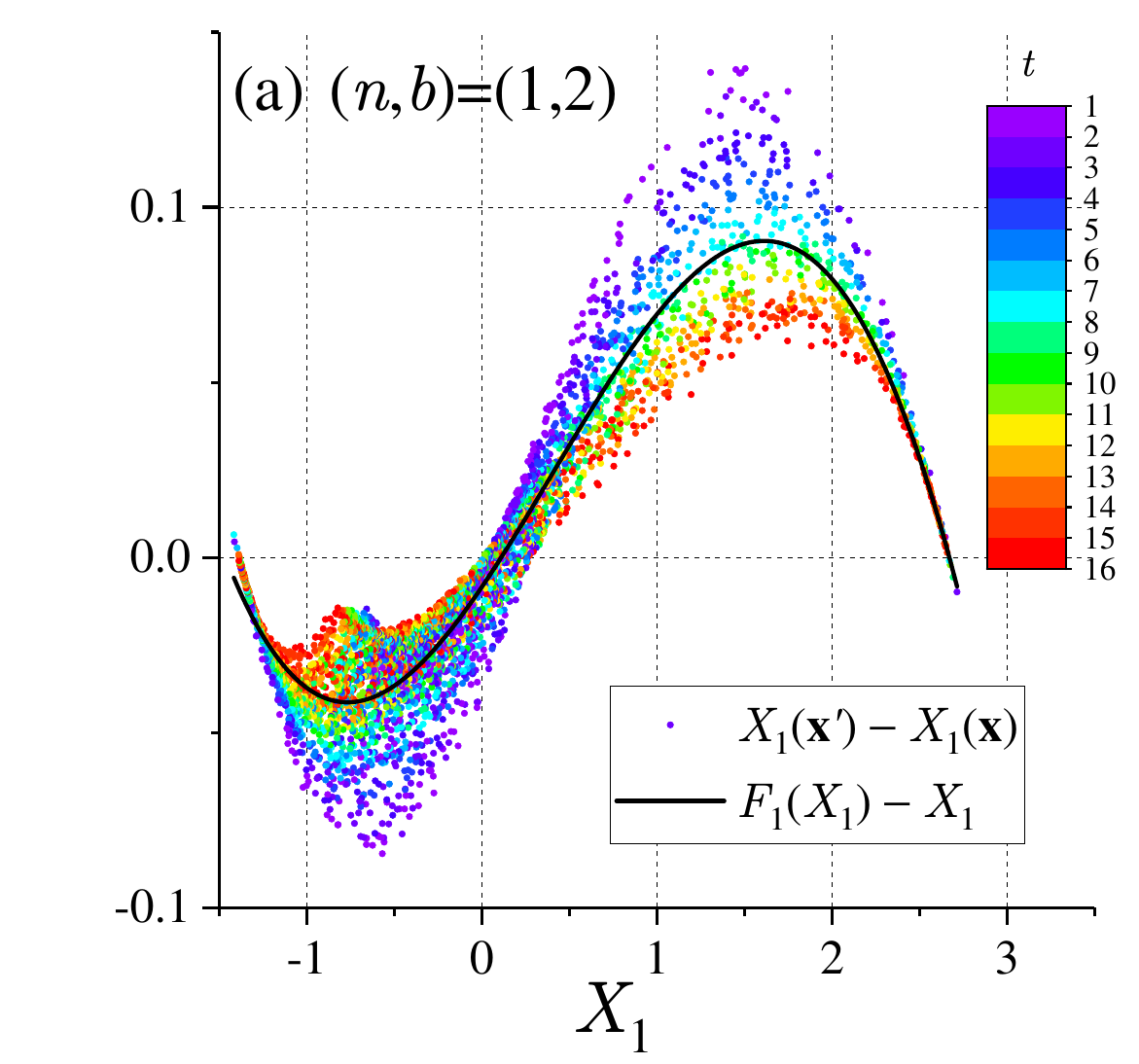}
    \includegraphics[trim=0 50 280 20,scale=0.35,clip]{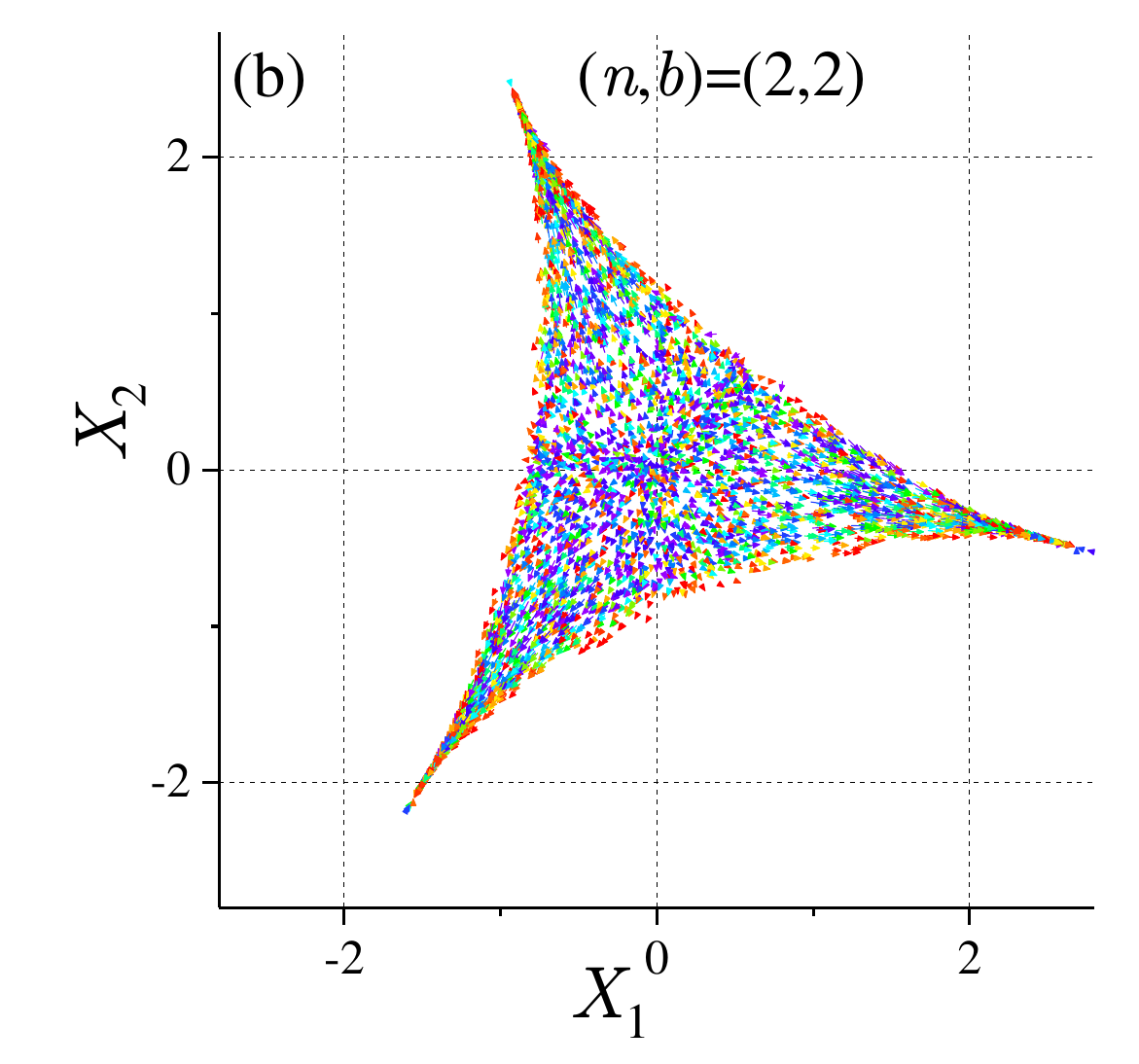}
    \\
	\hspace{15mm}
    \includegraphics[trim=160 200 380 80,scale=0.56,clip]{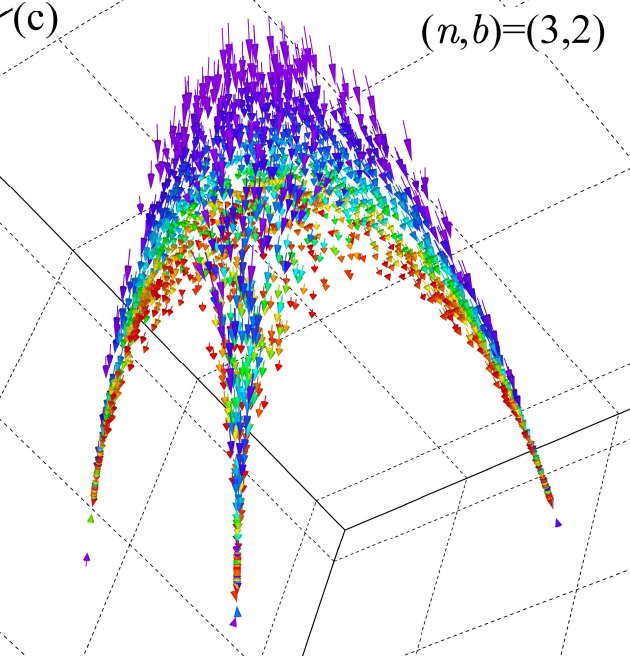}
    \includegraphics[trim=0 50 220 20,scale=0.35,clip]{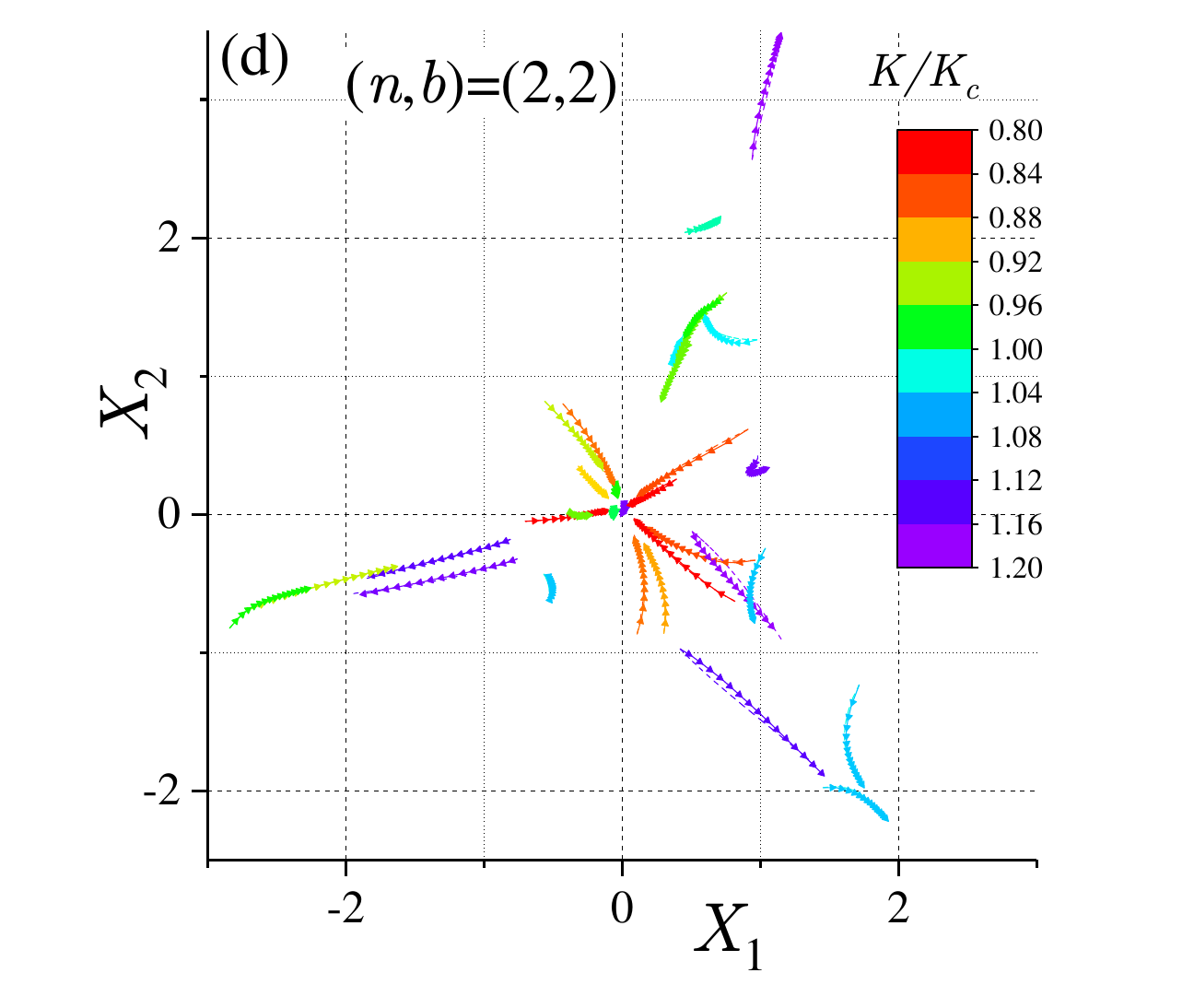}
\end{center}
    \caption{
        The displacement of the reduced macroscopic variables for IC I. 
The panels (a), (b), and (c) correspond to $n$=1, 2, and 3, respectively. 
The symbols and arrows are colored with respect to the observation time of the data. 
The coupling strength $K$ equals $1.2K_c$ except  panel (d), 
where $K$ distributes in $[0.8K_c, 1.2K_c)$.   
        \label{fig:IC1}
    }
\end{figure}

Figure~\ref{fig:IC1}(a) plots the one-step displacement of $X_1$ as a function of $X_1$ for $(n,b)=(1,2)$.  
The symbols show $X_1(\yy)-X_1(\xx)$ of the data points, 
which looks far from a single-valued function of $X_1$. 
The solid curve indicates $F_1(X_1)-X_1$, 
in which we can find an unstable fixed point (FP) at $X_1 \approx 0.1$ 
and stable FPs at $X_1\approx-1.4$ and $X_1\approx2.7$. 
These are speculated to correspond to the paramagnetic state and the distinct ferromagnetic states, respectively.  
The values of $X_1$ are related to $M_\sigma = 0, -M_\mrm{eq}/2, M_\mrm{eq}$, respectively,   
where $M_\mrm{eq}$ is the equilibrium value of the magnetization of the majority state.

In Fig.~\ref{fig:IC1}(b), the displacements $\XX(\xx) \to \XX(\yy)$ for $(n,b)=(2,2)$ are indicated by arrows. 
Three-fold symmetry is obvious and we can find a paramagnetic FP at the origin and three ferromagnetic FPs around it. 
Note that the LF is invariant against the rotation of $\XX$ in $n$-dimensional space. 
Since a similar flow structure is obtained for $b=1$, 
$\XX$ is presumably a quantity like the magnetization. 
Interestingly, the two-dimensional magnetization is preferred for $n=2$ 
rather than the pair of the one-dimensional magnetization and the internal energy.

In Fig.~\ref{fig:IC1}(c), the displacements for $(n,b)=(3,2)$ are indicated by arrows. 
It seems to be made by adding a vertical axis, corresponding to the internal energy, 
to the plain in Fig.~\ref{fig:IC1}(b).  
As mentioned previously, this considerably improves the precision of the macroscopic DS.

\subsection{initial condition II: $p_1=p_2$}

\begin{figure}[t]
    \includegraphics[trim=20 60 280 20,scale=0.35,clip]{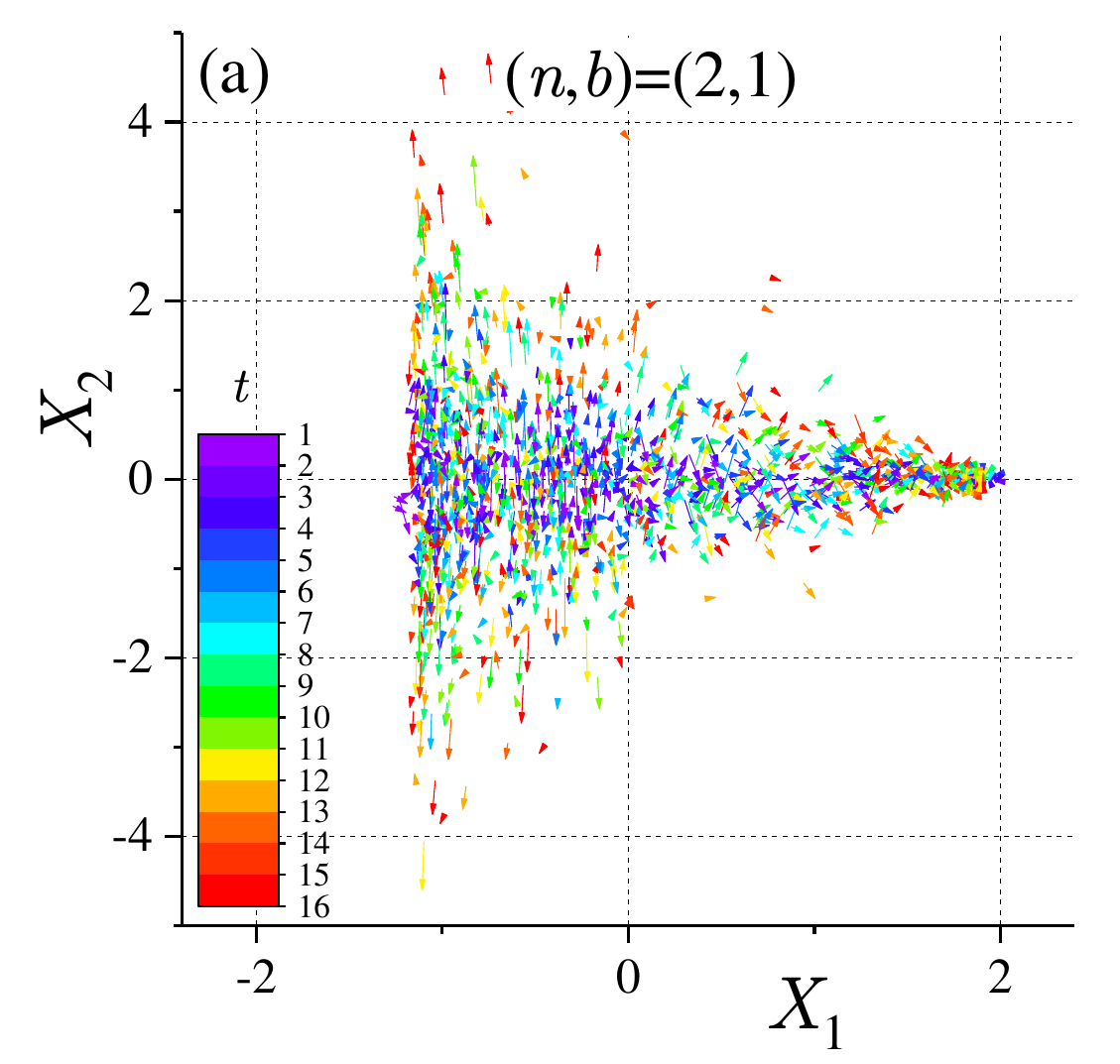}
    \includegraphics[trim=20 60 280 20,scale=0.35,clip]{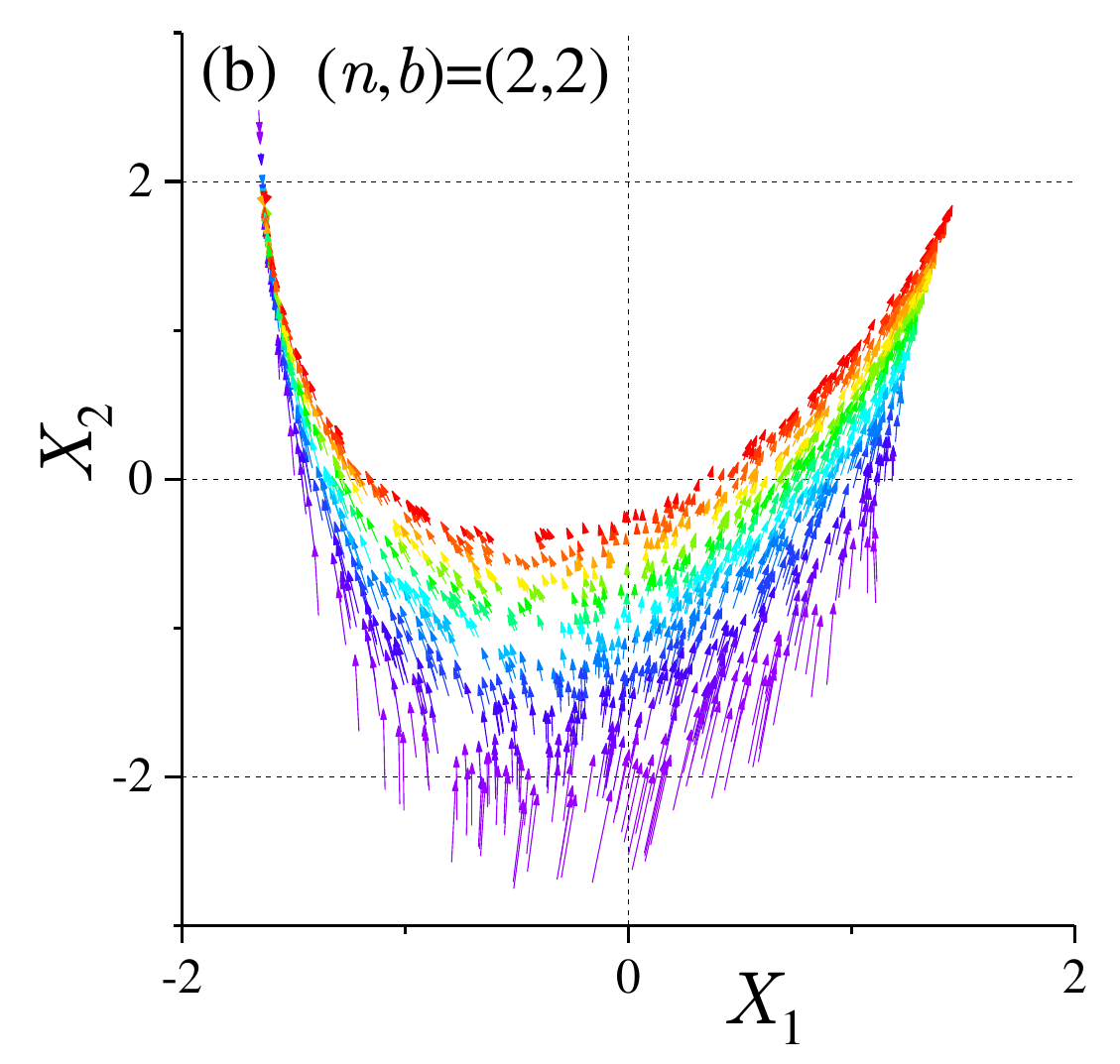}
	\\
    \hspace{15mm}
	\includegraphics[trim=270 150 240 130,scale=0.56,clip]{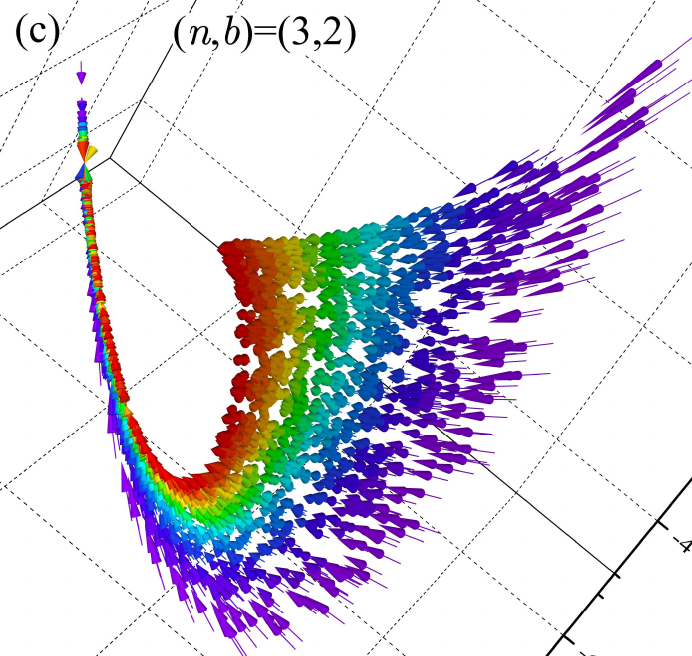}
    \caption{
        The displacement of the reduced macroscopic variables for IC II. 
        \label{fig:IC2}
    }
\end{figure}

Under IC II, the fraction of the states 1 and 2 are almost equal at $t=0$, 
namely, $\tx_1 \approx \tx_2 \approx (1-\tx_0)/2$.
If the state 0 is the majority at $t=0$, it remains so for $t>0$. 
Else, state 1 or 2 becomes the majority after stochastic symmetry-breaking 
but it takes a very long time. 
Figure~\ref{fig:L-b}(b) shows the $b$-dependence of $L_n^*$ for $n=1,\dots,4$. 
For $b \ge 2$, $L_n^*$ is the lowest at $n=2$ and the values are quite small. 
Contrastingly, $L_n^*$ is quite large for $(n,b)=(2,1)$. 

While we obtain similar results with that for IC I for $n=1$, 
this is not the case for $n=2$. 
Figure~\ref{fig:IC2}(a) shows the displacements for $(n,b)=(2,1)$, 
where $X_1$ and $X_2$ seems to correspond to $\tx_0$ and $\tx_1 - \tx_2$, respectively. 
We can find a stable FP at $(1.8,0.0)$, an unstable FP at $(-0.3,0.0)$ 
and a saddle point at $(-1.0, 0.0)$. 
It is speculated that there are two additional stable FPs at $(-1.0, \pm \infty)$. 
Since states 1 and 2 are almost balanced, 
$\tx_1 - \tx_2$ is not extensive, namely $O(N^{-1/2})$. 
Therefore, $X_2$ is in the same order as stochastic fluctuation. 
This is the reason that $L_n^*$ is quite large for $(n,b)=(2,1)$. 

In Fig.~\ref{fig:IC2}(b) for $(n,b)=(2,2)$, 
$X_1$ and $X_2$ seems to correspond to $M_0$ and $-E$, respectively. 
There are two stable FPs; one is at $(-1.6, 1.9)$ where state 0 is the majority  
and the other is at $(1.5, 1.9)$ where it is the minority. 
Figure~\ref{fig:IC2}(c) for $(n,b)=(3,2)$ seems to be made by 
bending Fig.~\ref{fig:IC2}(b) in a three dimensional space. 
This suggests that $\txx \in \RR^{N_b}$ is actually on a two-dimensional manifold. 
This is also the case for $b \ge 3$. 
These imply that $n=2$ is sufficient for the dynamics under IC II.

\subsection{initial condition III: $p_0=p_1=p_2$}

\begin{figure}[t]
    \includegraphics[trim=20 60 280 20,scale=0.35,clip]{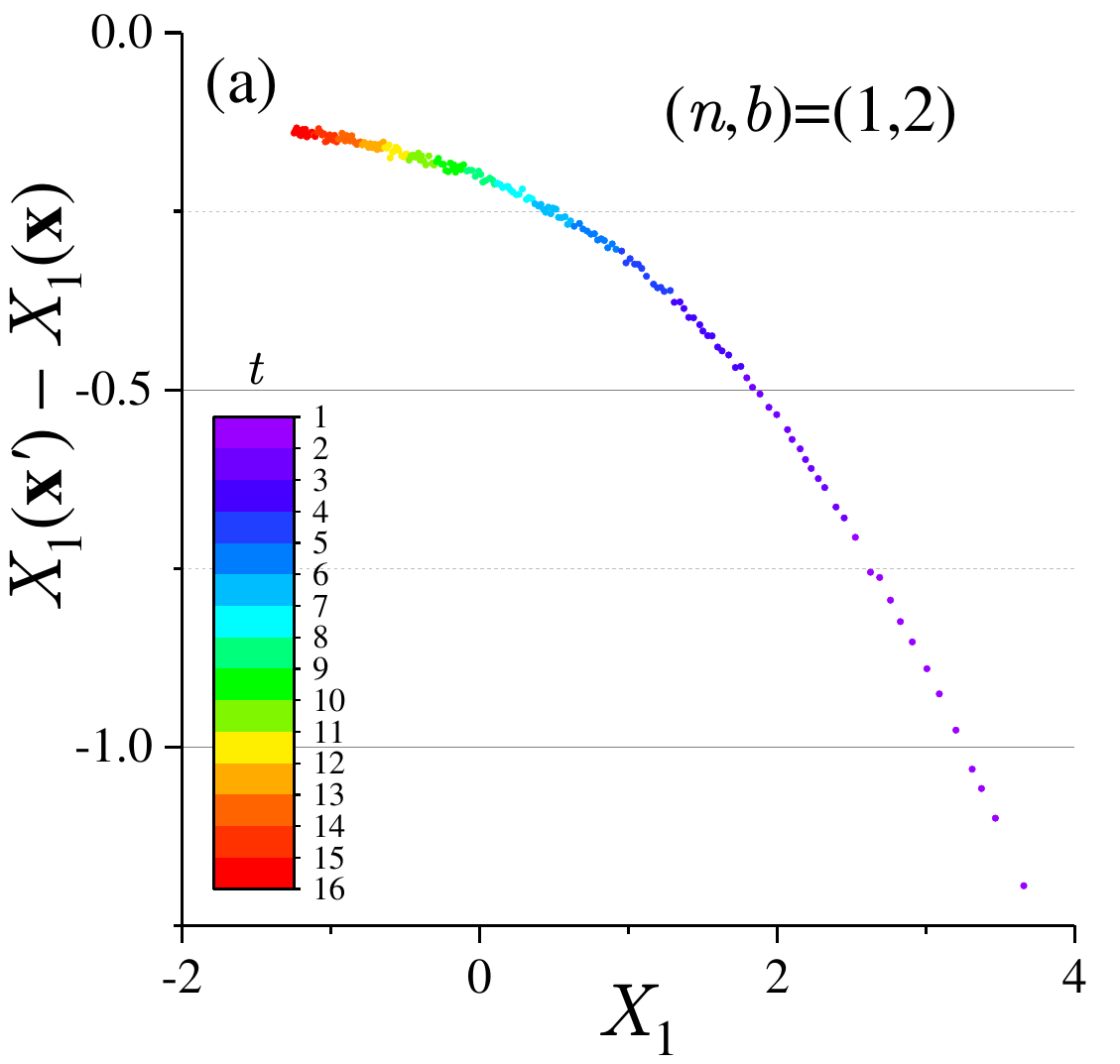}
    \includegraphics[trim=20 60 280 20,scale=0.35,clip]{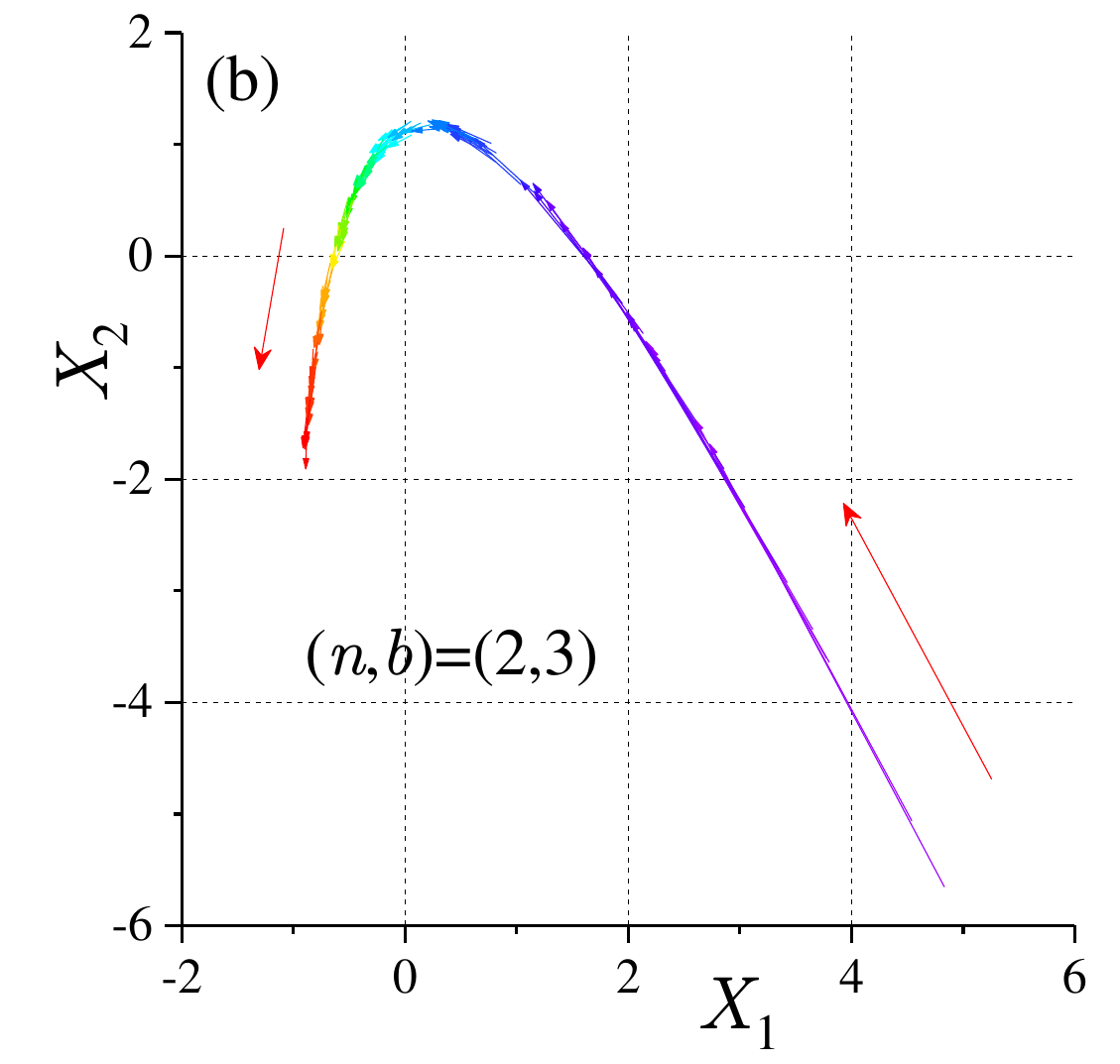}
    \caption{
        The displacement of reduced macroscopic variables for IC III. 
        \label{fig:IC3}
    }
\end{figure}

Under IC III, the initial state is macroscopically unique, 
namely, $\tx_0 \approx \tx_1 \approx \tx_2 \approx 1/3$ at $t=0$. 
Consequently, arbitrary extensive variable, such as internal energy, 
exhibits unique time evolution for $N\to\infty$ 
and, therefore, it trivially has a closed DS as far as it is monotonic with respect to $t$. 
For $b=1$, all components of magnetization are subextensive 
and we observe noisy diffusion of the magnetization both for $n=1$ and 2. 
Figure~\ref{fig:IC3}(a) plots the displacement of $X_1$ for $(n,b)=(1,2)$, 
which is obviously a single-valued function of $X_1$. 
Here, $X_1$ is presumably a quantity similar to $E$. 
As seen in Fig.~\ref{fig:L-b}(c), $L_n^*$ with $n=1$ decreases a lot as $b$ increases from 1 to 2. 
For $n=2$, a similar drop occurs as $b$ increases from 2 to 3. 
This is because it is impossible to make two linear-independent extensive variables for $b \le 2$. 
Figure~\ref{fig:IC3}(b) for $(n,b)=(2,3)$, however, obviously indicates that $X_1$ and $X_2$ has nonlinear dependence. 
These imply that $n=1$ is sufficient for IC III.

\subsection{data with distribution of the coupling strength}

Next, we show the results for the data with various $K$.
Each sample has different $K$ as $K \sim U(0.80K_c, 1.20K_c)$. 
Then, we need to replace $\FF(\XX)$ with $\FF(\XX, K)$. 
The implementation in learning is easy; 
we regard $K$ as the $(n+1)$-th variable in the polynomial in Eq.~\eqref{eq:polynomial_of_F}.

Figure~\ref{fig:IC1}(d) shows several trajectories of $\XX(\xx)$ 
driven by the microscopic DS $\ff$ with various $K$. 
While those for $K<K_c$ approach the paramagnetic FP at the center, 
those for $K>K_c$ approach one of the three ferromagnetic FPs, which move as $K$ changes.

Figure~\ref{fig:L-b_compare}(a) compares $L_n^*$ for $K \in [0.80K_c, 1.20K_c)$ 
and $L_n^*$ for unique $K(=1.20K_c)$. 
There are small differences between them, 
which implies that the macroscopic variable need not be optimized specifically for each $K$. 
For small $n$ and $b$, however, $L_n^*$ for various $K$ is smaller than that for $K=1.20K_c$. 
This is because the flow structure in the space of $\XX$ is simpler for $K<K_c$ than for $K>K_c$ 
and, therefore, the loss function is smaller for smaller $K$. 
For large $n$ and $b$, we find the opposite tendency: the former is larger than the latter. 
This is presumably because the incompleteness of the learning is enhanced by the increase of the elements of $\WW$.

\subsection{dropout}
\label{subsec:dropout}

\begin{figure}[t]
    \includegraphics[trim=10 60 420 20,scale=0.35,clip]{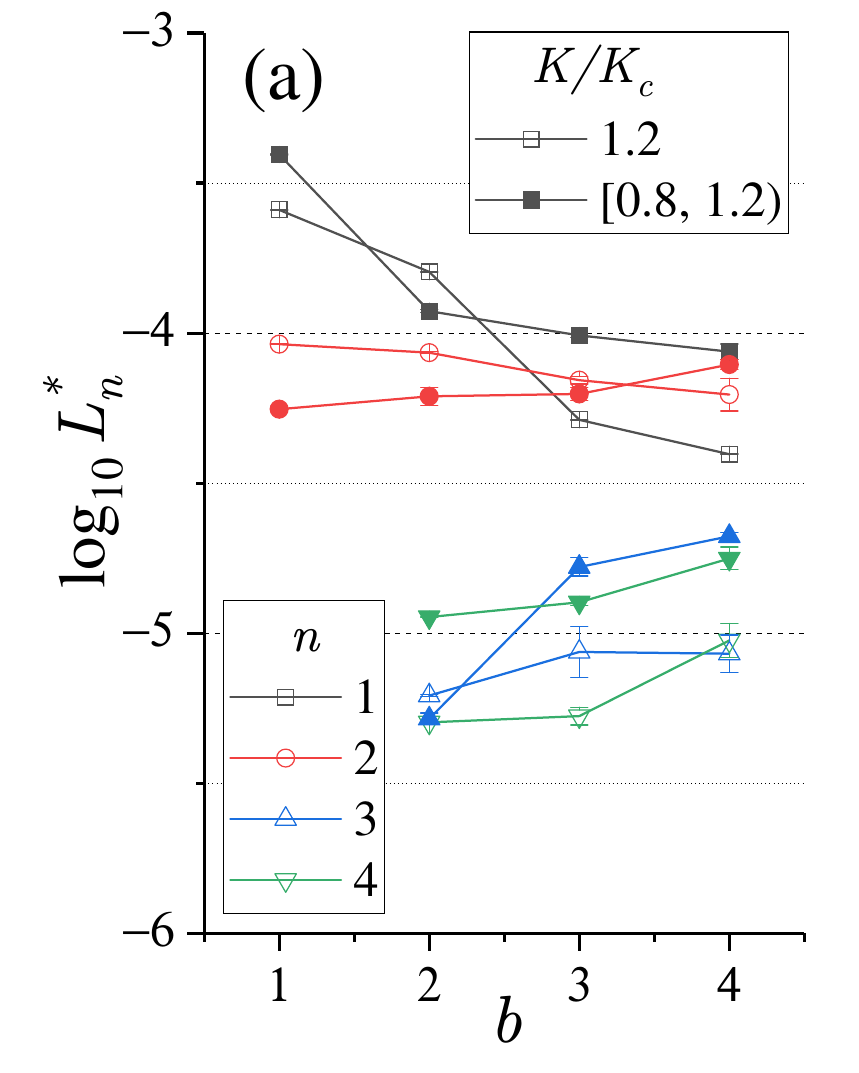}
    \includegraphics[trim=10 60 420 20,scale=0.35,clip]{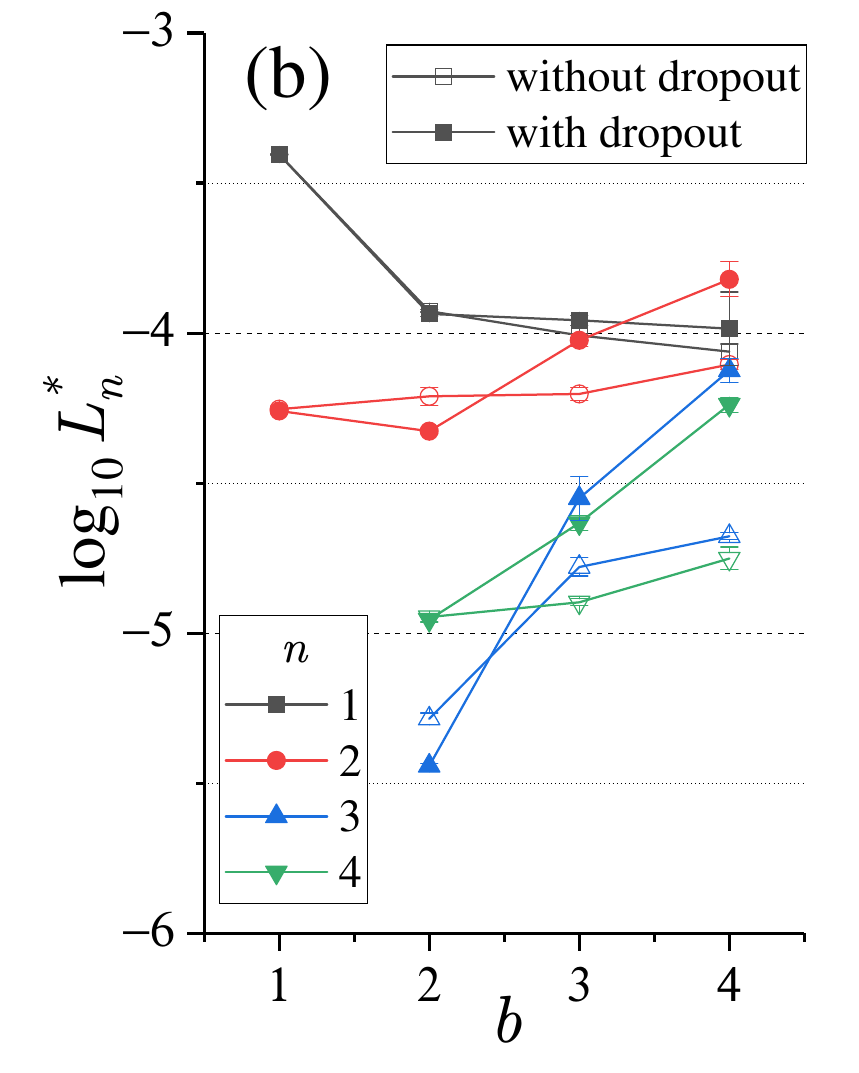}
    \includegraphics[trim=10 60 420 20,scale=0.35,clip]{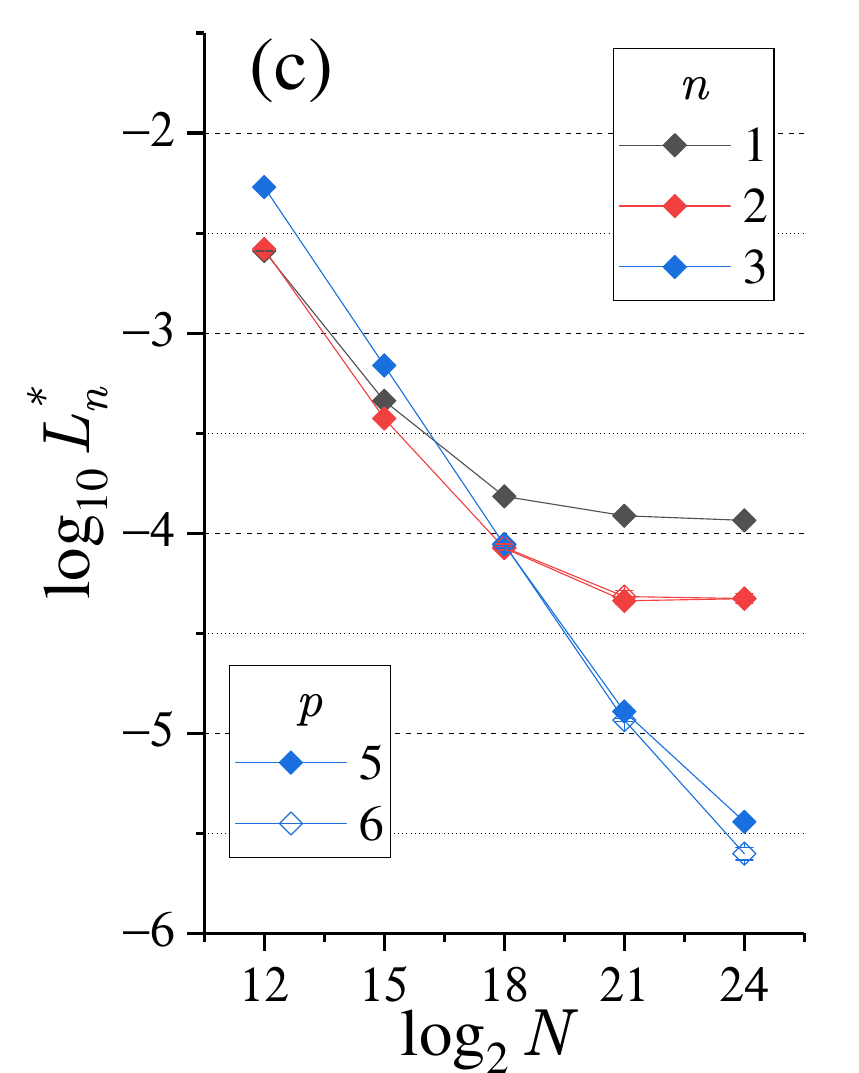}
    \caption{
        (a) Comparison of $L_n^*$ for various $K$ and that for unique $K$. 
        (b) Comparison of $L_n^*$ with dropout and without dropout. 
        (c) System-size dependence of $L_n^*$.
The results of all panels are for IC I.  
        \label{fig:L-b_compare}
    }
\end{figure}

\begin{figure}[t]
    \includegraphics[trim=10 60 420 20,scale=0.35,clip]{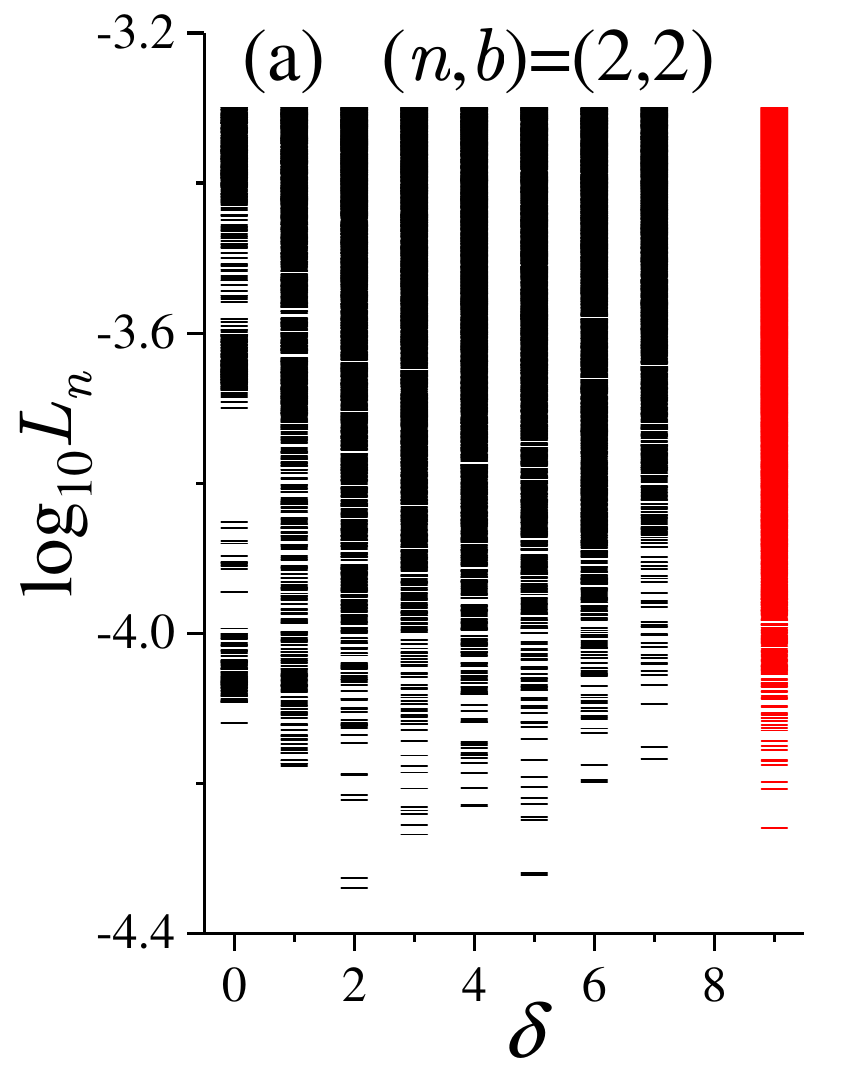}
    \includegraphics[trim=10 60 420 20,scale=0.35,clip]{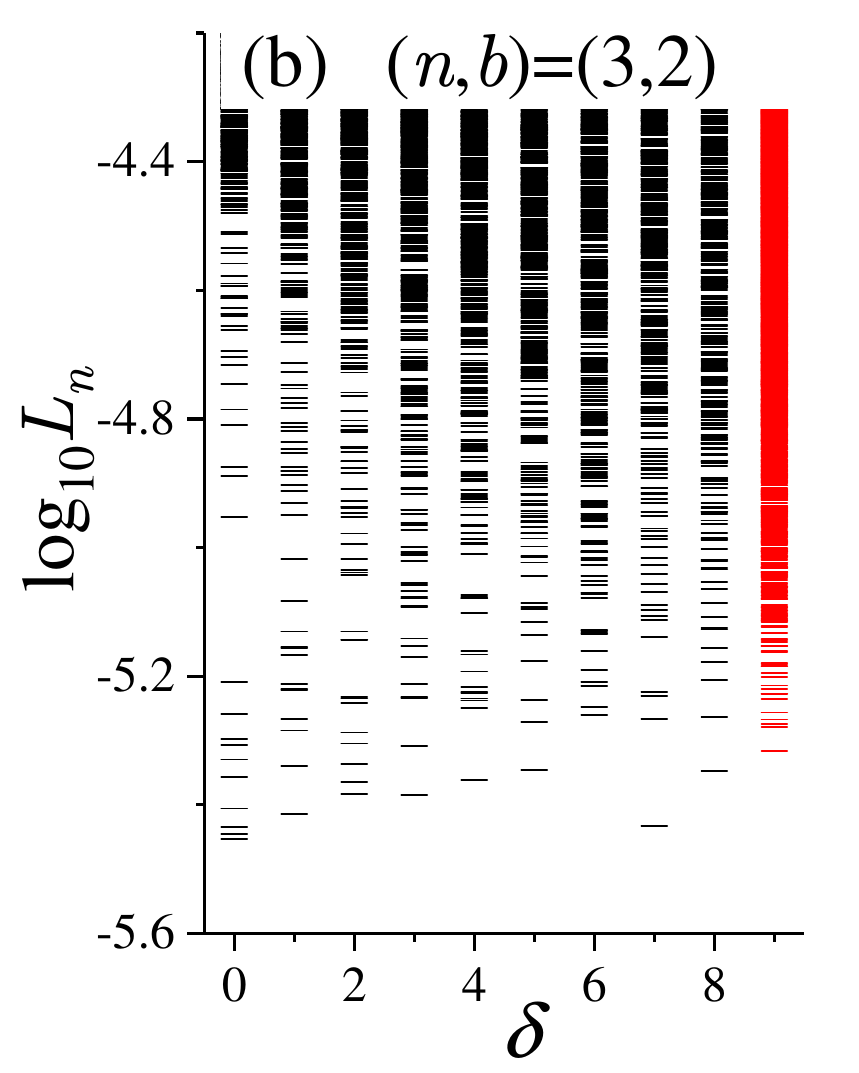}
    \includegraphics[trim=10 60 420 20,scale=0.35,clip]{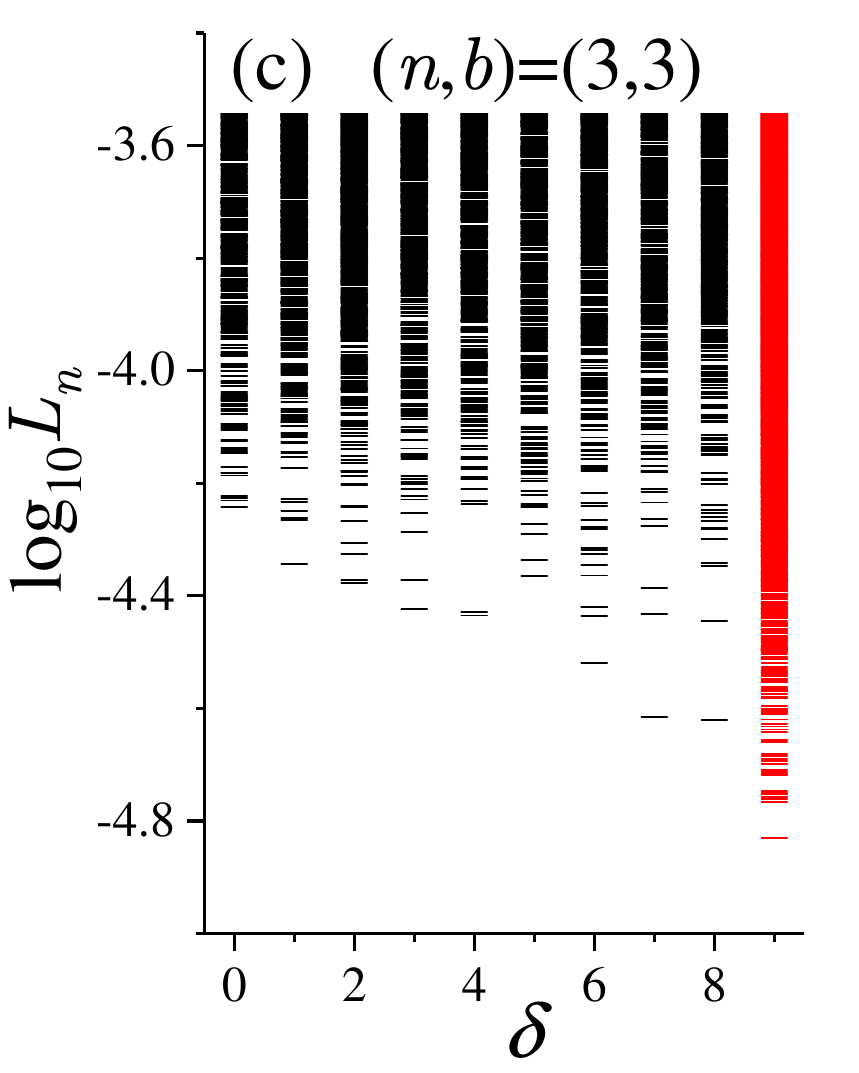}
    \caption{
Distribution of local minima of $L_n$ as a function of redundancy of the elements of $\ww$. 
These values are recorded when we temporally abandon the learning and initialize $\ww$.   
The most right column in each panel shows the results without dropout. 
        \label{fig:L-delta}
    }
\end{figure}

Figure~\ref{fig:L-b_compare}(b) compares $L_n^*$ with dropout and $L_n^*$ without dropout. 
For $b \ge 3$, dropout makes $L_n^*$ larger. 
This is natural because dropout reduces the ability to the expression of $\XX$. 
On the other hand, dropout makes $L_n^*$ a little smaller for $b=2$. 
This is because the learning becomes more efficient  
by the reduction of the number of tuning parameters.
Figure~\ref{fig:L-delta} plots the local minimum values of $L_n$ obtained by the learning as a function of $\delta$. 
As seen in the panel (a) and (b) for $b=2$, 
the global minimum is obtained with relatively small $\delta$. 
Remarkably, the best result is obtained with $\delta=0$ for $(n,b)=(3,2)$. 
For $(n,b)=(2,3)$, the best solution is obtained with $\delta=2$. 
The lower bounds for $\delta < 2$ are considerably larger than that for $\delta=2$. 
For $(n,b)=(3,3)$, contrastingly, 
the lower bound of $L_n$ tends to decrease with $\delta$ as far as $\delta \le 8$. 

Next, let us watch the expression of $\XX$. 
For $(n,b)=(3,2)$, the best two solutions are 
$( \tx_{11}, \, \tx_{22} \!-\! 0.44 \tx_{11}, \, \tx_{00} \!-\! 0.80 \tx_{11} \!-\! 0.80 \tx_{22})$ 
and 
$( \tx_{11}, \, \tx_{00} \!-\! 0.44 \tx_{11}, \, \tx_{22} \!-\! 0.80 \tx_{00} \!-\! 0.79 \tx_{11})$. 
(Here, we normalized each component of $\XX$ so that the largest coefficient becomes one.)
These are nearly equivalent after the permutation of states 0 and 2 
(Note that there is no degree of freedom in $\ww$ for $\delta=0$). 
It is remarkable that only the bases with the same states, such as $\tx_{00}$ and $\tx_{11}$, are employed. 
This may be because such bases tend to change monotonically with time 
and the aggregation of the trajectories forms a directed manifold. 
For $(n,b)=(2,3)$, the two best solutions  are 
$( \tx_{00} \!-\! 0.68\tx_{11}, \, \tx_{22} \!-\! 0.46 \tx_{00} \!-\! 0.65 \tx_{11})$ 
and 
$( \tx_{11} \!-\! 0.74\tx_{00}, \, \tx_{22} \!-\! 0.41 \tx_{00} \!-\! 0.27 \tx_{11})$. 
Again, the bases with different states do not appear.

\subsection{large-size limit}

The LF contains the systematic error coming from the finiteness of the system size. 
Here we consider the large-$N$ limit. 
Figure~\ref{fig:L-b_compare}(c) shows the $N$-dependence of $L_n^*$ for $b=2$. 
As $N$ increases, $L_n^*$ looks to converge to a finite value for $n=1$ and 2.  
On the other hand, $L_n^*$ keeps decreasing to zero as a power function of $N$ for $n=3$, 
which suggests that rigorously closed DS exists for $N \to \infty$. 
Although the slope becomes a little gentler between $2^{21}$ and $2^{24}$,  
this is presumably due to the disability of expression of $\FF$. 
Actually, $L_n^*$ for $N=2^{24}$ is decreased by the increment of the order of the polynomial from 5 to 6.  
Under IC II and III, $L_n^*$ similarly decreases to zero for $n=2$ and $n=1$, respectively (not shown).

\subsection{data-size dependence of learning}

\begin{figure}[t]
\includegraphics[trim=10 350 240 0,scale=0.875,clip]{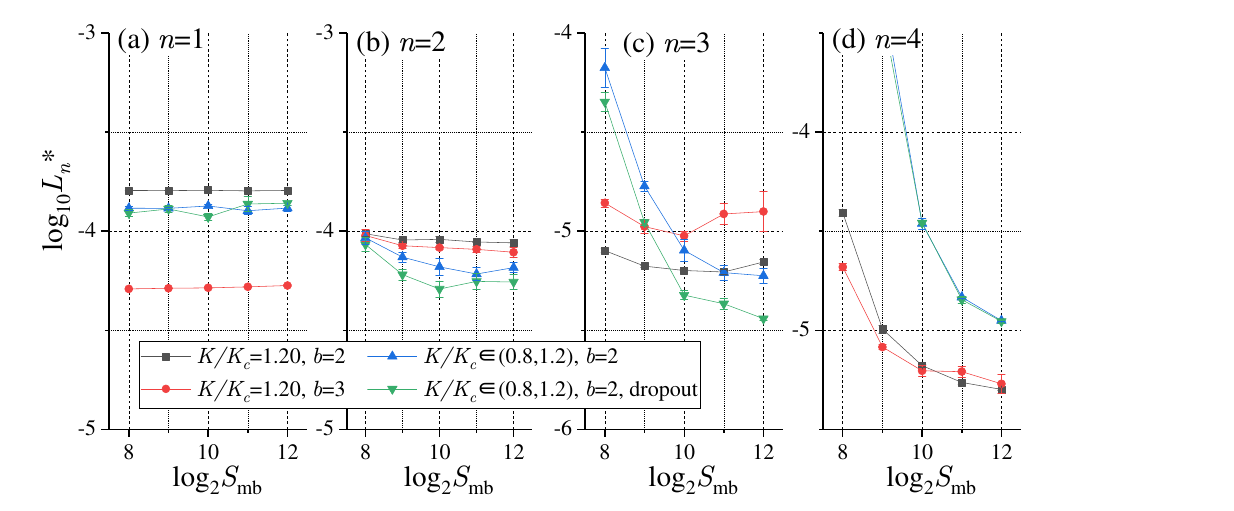}
\caption{
The average of the best four optimal values of $L_n$ for each minibatch size $S_mb$ is plotted. 
\label{fig:L-Smb}
}
\end{figure}

Here we consider the interplay between the efficiency of the learning and the data size. 
Figure \ref{fig:L-Smb} shows the optimum value of $L_n$ 
that is obtained by the learning with the minibatch size $\Nmb$. 
Four panels corresponds to $n=1, \dots, 4$, respectively.  
Each panel contains four graphs that differ in terms of $b$, distribution of $K$, and dropout. 
For $n=1$, we find little $\Nmb$-dependence, which means even the minimum size $2^8$ is sufficient. 
For $n\ge2$, there are certain thresholds in $\Nmb$ below which $L_n^*$ increases with decreasing $\Nmb$. 
The thresholds tend to increase with $n$. 
They are also increased by introducing the distribution of $K$     
but rarely affected by $b$ and dropout. 
These imply that the dimension of the argument of $\FF$ raises the threshold. 
The number of the polynomial bases for $\FF$ equals 6, 21, 56, 126, 252 for $n=1, \dots, 5$, respectively. 
In addition, the robust orthogonalization of $\XX$ that satisfies 
the acceptance condition Eq.~\eqref{eq:cond_orthonormalization}, requires more data as $n$ increases.

Unfortunately, the present machine learning does not seem to benefit from using mini-batches because $L_n^*$ decreases almost monotonically with $\Nmb$. 
This is presumably because a considerably high-precision regression is required for $\FF$ for small $L_n^*$.

\section{Summary and Discussions}

In this study, a comprehensive framework for the dimensional reduction of DSs by machine learning is proposed. 
In particular, the implementation of the method to the system with discrete and bounded variables is considered, 
and the application to the three-state Potts model is demonstrated. 
The obtained macroscopic DS exhibits plausible behavior. 
It is robust against the distribution of the value of the coupling strength. 
It was preliminarily confirmed that it is also robust against the imposition of distributed external magnetic fields. 
Successful cases are also found, where the dropout of the matrix elements that characterize the reduced variable does not raise the loss function. 
Furthermore, a consequence that $\lim_{N \to \infty} L_n^*$ equals zero above a certain threshold $n=n_c$, which depends on IC, is obtained; 
$n_c$ decreases as the symmetry of the initial state increases. 
For $n>n_c$, $\XX(\xx)$ are embedded on an $n_c$-dimensional manifold.

Almost closed DS is obtained with $b=2$. 
This is reasonable because the model analyzed has only nearest-neighbor interaction.  
Although $L_n^*$ decreases with $b$ above 2, 
this is not considered to be essential. 
The increment of $b$ raises the degree of freedom to modify the distribution of $\XX(\yy)$ in $n$-dimensional space. 
This reduces $L_n$ to some extent 
but the flow structure of $\XX$, which is represented by the configuration of FPs, does not change. 
In the same sense, the comparison of $L_n^*$'s for different $n$ may not be meaningful. 
It would be significant to consider a more proper loss function. 

The proposed method is expected to work for most DSs with discrete variables, such as cellular automaton and contact process. 
Developing feasible implementations for off-lattice systems and systems with continuous degrees of freedom is challenging. 
The proposed method has several additional uses. 
A prediction formula for the quantity of interest, such as an order parameter, can be found by fixing a part of the components of $\XX$. 
In addition, a conserved quantity can be found by allowing $\FF$ identity transformation.

\section*{Acknowledgments}
This work was supported by JSPS KAKENHI Grant Number JP18K03469.

\section*{References}



\providecommand{\newblock}{}

\end{document}